\documentclass[10pt,prl,aps,twocolumn,showpacs,floatfix,superscriptaddress]{revtex4-1}
\usepackage{graphicx}
\usepackage{amssymb}
\usepackage{color}
\usepackage{amsmath}  
\newcommand{\be}{\begin{equation}}
\newcommand{\ee}{\end{equation}}
\newcommand{\ba}{\begin{eqnarray}}
\newcommand{\ea}{\end{eqnarray}}

\newcommand{\gae}
{\,\hbox{\lower0.5ex\hbox{$\sim$}\llap{\raise0.5ex\hbox{$>$}}}\,}
\newcommand{\lae}
{\,\hbox{\lower0.5ex\hbox{$\sim$}\llap{\raise0.5ex\hbox{$<$}}}\,}

\begin{document}

\title{Anomalous quantum glass of bosons in a random potential in two dimensions}

\author{Yancheng Wang}
\affiliation{Department of Physics, Beijing Normal University, Beijing 100875, China}
\affiliation{Institute of Physics, Chinese Academy of Sciences, Beijing 100190, China}

\author{Wenan Guo}
\email[e-mail: ]{waguo@bnu.edu.cn}
\affiliation{Department of Physics, Beijing Normal University, Beijing 100875, China}
\affiliation{State Key Laboratory of Theoretical Physics, Institute of Theoretical
Physics, Chinese Academy of Sciences, Beijing 100190, China}

\author{Anders W. Sandvik}
\email[e-mail: ]{sandvik@bu.edu}
\affiliation{Department of Physics, Boston University, 590 Commonwealth Avenue, Boston, Massachusetts 02215, USA}

\begin{abstract}
We present a quantum Monte Carlo study of the ``quantum glass'' phase of the 2D Bose-Hubbard model with random potentials
at filling $\rho=1$. In the narrow region between the Mott and superfluid phases the compressibility has the form 
$\kappa \sim {\rm exp}(-b/T^\alpha)+c$ with $\alpha <1$ and $c$ vanishing or very small. Thus, at $T=0$ the system is 
either incompressible (a Mott glass) or nearly incompressible (a Mott-glass-like anomalous Bose glass). At stronger disorder, 
where a glass reappears from the superfluid, we find a conventional highly compressible Bose glass. On a path connecting these states, away 
from the superfluid at larger Hubbard repulsion, a change of the disorder strength by only $10\%$ changes the low-temperature 
compressibility by more than four orders of magnitude, lending support to two types of glass states separated by a phase
transition or a sharp cross-over.
\end{abstract}

\date{\today}


\maketitle

There are two types of ground states of interacting lattice bosons in the absence of disorder; the superfluid (SF) and the Mott-insulator (MI). 
In the Bose-Hubbard model (BHM) with repulsive on-site interactions \cite{giamarchi88,fisher89} an MI state has an integer number 
of particles per site and there is a gap to states with added or removed particles. The gapless SF can have any filling fraction. These phases and the 
quantum phase transitions between them are well understood \cite{giamarchi88,fisher89,freericks96,batrouni02,capogrosso08,mahmud11} and have been 
realized experimentally with ultracold atoms in optical lattices \cite{greiner02,bloch08}. 

If disorder in the form of random site potentials is introduced in the BHM (which can also be accomplished in optical lattices \cite{damski03,fallani07})
a third state appears---an insulating but gapless quantum glass. This state has been the subject of numerous studies 
\cite{giamarchi88,fisher89,freericks96,krauth91,wallin94,giamarchi01,lee01,pazmandi98,wu08,alet03,prokofev04,sengupta07,weichman08,pollet09,gurarie09,kruger11,lin11,soyler11,altman04,altman10}
but many of its properties are still not well understood. Two types of glass states are known; the compressible Bose glass (BG) and the incompressible Mott glass (MG), 
with the latter commonly believed to appear only at commensurate filling fractions in systems with particle-hole 
symmetry \cite{prokofev04,sengupta07,weichman08,altman04,altman10,roscilde07,iyer12}. The currently prevailing notion is that the glass state in the 
2D BHM with random potentials is always of the compressible BG type \cite{weichman08,pollet09,gurarie09,soyler11}. 

We here present quantum Monte Carlo (QMC) results for the two-dimensional (2D) site-disordered BHM, showing that there is actually 
an extended parameter region in which the BG is either replaced by an MG or has an anomalously small (in practice undetectable) compressibility. The system
is described by the Hamiltonian
\begin{equation}
H = - t \sum_{\langle ij\rangle} (b^+_ib_j + b^+_jb_i) 
 + \frac{U}{2}\sum_{i=1}^N n_i(n_i-1)  + \sum_{i=1}^N \epsilon_i n_i, 
\label{hamiltonian}
\end{equation}
where $\langle ij\rangle$ are nearest neighbors on the square lattice, $b^+_i$ ($b_i$) are boson creation (destruction) operators, 
$n_i=b^+_ib_i$ site occupation numbers, and $\epsilon_i$ random potentials uniformly distributed in the range $[-\Lambda-\mu,\Lambda-\mu]$ about 
the average chemical potential $\mu$. We study the model using the stochastic series expansion (SSE) QMC method with directed loop updates \cite{sse}. 
We adjust the chemical potential so that the mean filling-fraction $\rho=\langle n_i\rangle=1$ (to within $<10^{-5}$) when averaged over sites 
$i$, disorder realizations, quantum and thermal fluctuations. To speed up the simulations, we impose a 
cut-off $n_i \le 2$ (some times $n_i \le 3$) which does not change the nature of the states. 
We study sufficiently large inverse temperatures $\beta=t/T$ and lattice sizes $L$ ($N=L^2$ sites) to address the ground state in 
the thermodynamic limit.

\begin{figure}
\center{\includegraphics[width=6.5cm, clip]{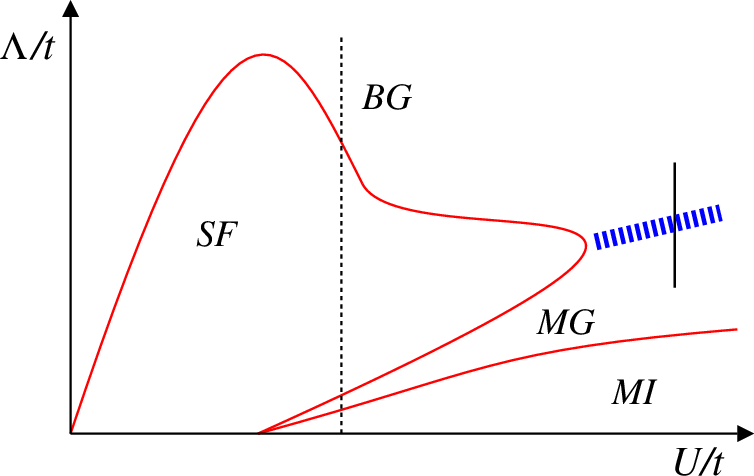}}
\caption{(Color online) Schematic phase diagram in the plane of repulsion $U$ and disorder strength $\Lambda$ of the 2D BHM with random potentials and
filling $\rho=1$. We study the system along the dashed line spanning the Mott insulator (MI), putative Mott glass (MG), superfluid (SF), and Bose glass (BG). 
The boundary or cross-over between the MG and BG is indicated by the thick dashed line, and we also traverse it in calculations along a vertical (solid) line.}
\vskip-2mm
\label{pd}
\end{figure}

In the plane $(\mu/U,t/U)$, for fixed disorder strength $\Lambda$, there are characteristic ``Mott lobes'' inside which the filling is integer, 
while outside $\rho$ changes with $\mu,U$ 
\cite{fisher89,freericks96,giamarchi01,krauth91,wallin94,lee01,pazmandi98,wu08,alet03,prokofev04,pollet09,gurarie09,kruger11,lin11,soyler11}.
The lobes are surrounded by a quantum glass (Griffiths) phase  for any $\Lambda > 0$ \cite{pollet09,gurarie09,kruger11,lin11,soyler11}. 
At fixed integer filling, in the plane $(U,\Lambda)$ there is a narrow ``finger'' of the glass phase intervening between the MI and SF, 
shrinking to a point $U_c$ at $\Lambda=0$. Most studies have focused on $\rho=1$ and the phase diagram in this case is qualitatively 
very similar in two \cite{soyler11} and three \cite{gurarie09} dimensions. We show a schematic phase diagram in Fig.~\ref{pd}. 

We focus first on a vertical line in the phase diagram in Fig.~\ref{pd} at moderate $U$. With increasing $\Lambda$ we can go from the MI, 
transition into the glass state in the finger region, then into the SF, and finally re-enter a glass state at much larger $\Lambda$. We also consider a line
to the right of the SF phase in Fig.~\ref{pd}, studying the evolution of the glass with increasing disorder when the SF is not crossed but 
(as we will show) the properties change dramatically.

\begin{figure}
\center{\includegraphics[width=7cm, clip]{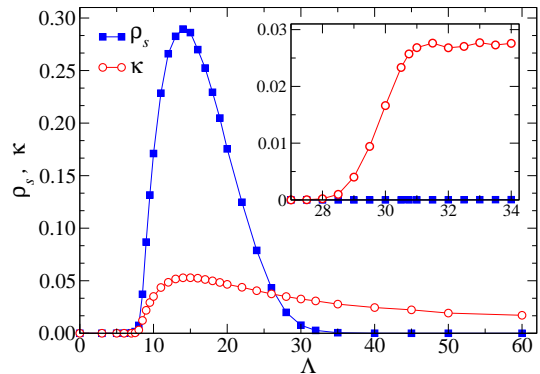}}
\caption{(Color online) Compressibility and superfluid stiffness vs disorder strength in a $16\times 16$ system at $\beta=8$. 
At $U=22$ (main graph), the system evolves from glass to SF and back to glass. Results for $U=60$ are shown in the inset. Here the 
system stays in the glass phase but the compressibility changes rapidly around $\Lambda=30$ (far away from the Mott boundary at $\Lambda \approx 24$).}
\label{l16}
\end{figure}

We compute two observables characterizing the states: the compressibility $\kappa$ and the superfluid stiffness $\rho_s$ (obtained with SSE
using, respectively, particle-number and winding number fluctuations \cite{sse}). The results were averaged over $500-1000$ realizations of
the random potentials. Fig.~\ref{l16} shows the evolution with $\Lambda$ of these quantities for a fixed system size at a low temperature. 

Before discussing the results, we recall the reasons for the existence of a glass phase and its expected nature.
In disordered systems in general, one can expect Griffiths phases where statistically rare large regions of some phase inside another 
phase lead to singularities not present in the absence of disorder \cite{griffiths69,mccoy69,vojta06}. For the integer-$\rho$ BHM with site disorder, the 
Griffiths argument states the following \cite{fisher89,freericks96}: Once the width $2\Lambda$ of the disorder distribution exceeds the Mott gap $\Delta_M$, 
there can be arbitrarily large domains of SF inside the MI. Until $\Lambda$ exceeds some larger critical value these domains are not 
percolating through the lattice, and the state is therefore insulating \cite{niederle13}. In the standard scenario (discussed further in
supplementary material), fluctuations of the overall chemical potential within the SF domains lead to near degeneracies of different particle-number 
sectors and, therefore, nonzero compressibility (a BG) \cite{pollet09,gurarie09,soyler11}. With these notions in mind, we now discuss our results.

Looking first at the superfluid stiffness in Fig.~\ref{l16}, the sharp increase at $\Lambda \approx 8$ 
signals the entry into the SF phase. A finite-size scaling analysis, presented below, shows that the transition takes place at 
$\Lambda_c \approx 8.3$, which is in reasonable  agreement with the result by S\"oyler {\it et al.}, $\Lambda_c(U=22)\approx 7.8$ \cite{soyler11}. 
Here we note that our model is slightly different, because of the cut-off $n_i\le 2$ (while there was no cut-off in Ref.~\cite{soyler11}). 
When we increase the cut-off to $n_i \le 3$ the critical point moves to a value consistent with
that of S\"oyler {\it et al}. Increasing $\Lambda$ further in Fig.~\ref{l16}, the superfluid stiffness eventually 
again decreases to zero at $\Lambda \approx 30$. This transition point is much smaller than that of S\"oyler {\it et al.}, $\Lambda_c\approx 70$, 
as would be expected in this region where the probability of site occupations beyond our cut-off is substantial.
Since the cut-off does not alter any symmetries of the system there is no reason to believe that it will affect 
our conclusions regarding the nature of the phases and transitions. 

\begin{figure}
\center{\includegraphics[width=7.5cm, clip]{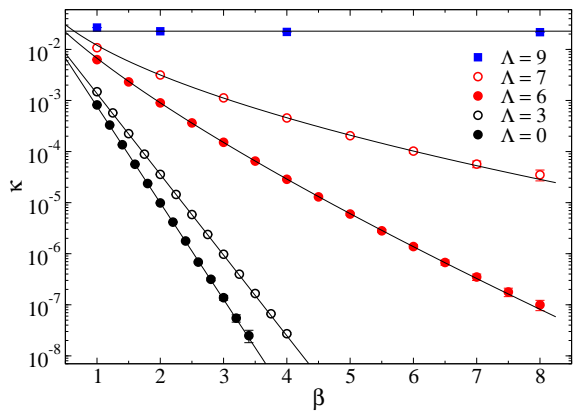}}
\caption{(Color online) Compressibility versus inverse temperature for $U=22$ and different 
strengths $\Lambda$ of the disorder. The lattice size $L=32$ in all cases, which is sufficient to eliminate finite-size effects at these temperatures. 
The curves are fits to the form (\ref{kappaform}) with $\alpha=1$ for $\Lambda =0,3$ (MI state) and $\alpha=0.78$ and $0.53$ for 
$\Lambda=6$ and $7$, respectively (MG state). For $\Lambda=9$ (SF state) $\kappa$ is essentially constant.}
\label{kappa}
\vskip-2mm
\end{figure}

Turning to the compressibility, it is substantial in the SF phase and when the glass is re-entered at large $\Lambda$.
However, it is very small below the SF transition, not only in the Mott phase (which extends up to $\Lambda \approx 4.3$ in our system,
based on the Mott gap of the clean MI as discussed in supplementary material) but also in the region $\Lambda \approx 5-7$, 
where the system is in a glass phase. The $L\to \infty$ compressibility as a function of temperature is 
shown in Fig.~\ref{kappa}. At $\Lambda=0$ and $3$, we observe the normal exponential decay with $\beta$ expected in the gapped MI phase. 
At $\Lambda=6$ and $7$ we instead find the form
\begin{equation}
\kappa \sim {\rm exp}(-b/T^\alpha)+c,
\label{kappaform}
\end{equation}
where $\alpha<1$ and $c=0$ (to within statistical errors). This form has
previously been found in random quantum spin systems \cite{roscilde07,ma14}, where $\kappa$ corresponds to the magnetic susceptibility and one 
expects it to vanish as $T\to 0$ because of spin-inversion symmetry (corresponding to particle-hole symmetry for bosons). 
Such an incompressible and insulating quantum glass is called an MG \cite{giamarchi01} and has also been shown to exist in variants of the 2D 
random BHM where particle-hole symmetry is explicitly built in \cite{altman10,iyer12} (and Ref.~\cite{giamarchi01} argued for its possible existence
also more generally). To our knowledge, $\kappa(T)$ was not computed for these systems and there is no theoretical prediction for its form. In the 
presence of random potentials there is no explicit particle-hole symmetry (but in principle there could be emergent particle-hole symmetry, as in the 
clean BHM at the tips of the Mott lobes \cite{fisher89}). It had been argued that the glass state of the BHM should then always be a clearly compressible 
BG \cite{pollet09,gurarie09,soyler11} (except very close to the MI boundary), contrary to our findings.

While $c=0$ in (\ref{kappaform}) may not hold strictly, the very small $\kappa(T \to 0)$ at the very least shows that the system is 
an anomalous BG, with exceedingly small (essentially undetectable) compressibility in a large part of the phase diagram. We here use the
term MG because, as we will argue below, even if $c>0$ but small the physics behind the anomalous BG is very similar to a true MG.

The form (\ref{kappaform}) of $\kappa(T)$ with $c=0$ can be understood heuristically as follows: Consider non-Mott domains 
below the percolation point inside an MI. At temperature $T$ there is some size $m$ such that for $T \alt m^{-a}$ all domains of size
$s<m$ are effectively in their ground states and do not contribute to the compressibility. The exponent $a$ depends on the low-energy level spectrum 
of the domains, which should be related to the fractal nature of the domains. Domains with $s>m$ should contribute essentially independently 
of $T$ and $s$. The probability of a site belonging to a domain of size $s>m$ is $\propto {\rm exp}(-dm^b)$, where $b$ can in principle be computed using classical percolation 
theory \cite{stauffer} (but may be different in a quantum system). In terms of the unknown exponents $a$ and $b$ the compressibility due to non-Mott 
domains is $\kappa \propto {\rm exp}(-dT^{-b/a})$, which is Eq.~(\ref{kappaform}) with $\alpha=b/a$. 

The above scenario neglects the arbitrarily close degeneracy of different particle-number sectors due to fluctuations of the average chemical
potential of the domains, which lead to $\kappa(T=0) > 0$ in the standard BG scenario (where the non-Mott domains are superfluid). How can these 
degeneracies be avoided? By studying isolated domains with a different chemical potential embedded in an MI, we have 
found (see supplementary metarial) that there are finite-size effects due to which particle-number degeneracies in the region of interest here only occur 
when the domains are large (with the critical size diverging at the Mott phase boundary). All domains below a critical size (which depends on the domain shape)
have vanishing $T \to 0$ compressibility and should not be regarded as superfluid---they are insulating because of finite size and effectively 
possess particle-hole symmetry at low energy. One still expects rare domains exceeding the critical size to contribute when $T \to 0$. However, we will show 
below that typical large domains should also have an altered spectral structure due to quantum-criticality when the SF boundary is approached. Thus, both small 
and large typical domains (the latter of which are fractals) may not contribute to the $T=0$ compressibility. 

Within the standard scenario, there should still exist rare large compressible domains in the Mott background,
but in reality the domains are never completely isolated from each other and the picture of degenerate single-domain levels 
may ultimately not be valid away from the atomic limit (large $U$ and $\Lambda$). Whether or not strictly $c=0$ in Eq.~(\ref{kappaform}), in practice the 
compressibility is undetectably small and the system is effectively an MG in the finger region (and, as we will see, also at larger $U$ in a substantial 
region along the Mott boundary).

\begin{figure}
\center{\includegraphics[width=7cm, clip]{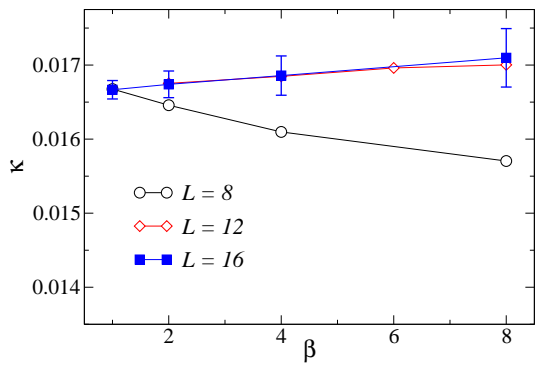}}
\caption{(Color online) Compressibility versus inverse temperature for systems with $U=22$, $\Lambda=60$.
Error bars are only shown for $L=16$ (being much smaller for $L=8,12$).}
\label{lam60}
\vskip-2mm
\end{figure}

We do find a compressible BG in the re-entrance region at large $\Lambda$ (above the SF in Fig.~\ref{pd}), as illustrated by results 
at $\Lambda=60$ in Fig.~\ref{lam60}. There should then be a phase transition or a cross-over separating the MG and the BG phases.
We have identified a dramatic  variation in the compressibility along a vertical line at  $U=60$. As shown in the inset of Fig.~\ref{l16}, at 
$\beta=8$, $\kappa$ increases rapidly with $\Lambda$ between $28$ to $31$ (which is far away from the Mott boundary at $\Lambda \approx 24$), 
before flattening out. The enhancement is more than four orders of magnitude, $\kappa$ being very small before the sharp increase. We do not find any significant 
finite-size effects in this region for $L>8$, and also the behavior does not change substantially upon further increasing $\beta$ (and
the $n_i$ cut-off also does not play a role here). The behavior therefore indicates a sharp cross over, not a phase transition, though
in principle $\kappa$ could still vanish exponentially at some point away from the Mott boundary. As a function of $U$, the cross-over most likely occurs 
on a line extending out from the right-side SF tip (``nose'') \cite{niederle13} in Fig.~\ref{pd} and can be interpreted as a change from a state where typical non-Mott
domains are not superfluid to a BG where the domains are superfluid but do not form a coherent global state.

We next study the critical $T=0$ compressibility at the lower glass--SF boundary, where $\kappa \sim (\Lambda - \Lambda_c)^{\nu(2-z)}$ is expected in the thermodynamic 
limit. If $z=2$, as is often assumed \cite{pollet09,gurarie09,soyler11}, $\kappa > 0$ is non-singular at the 
transition. One would then expect $\kappa >0$ also close to the transition inside the glass \cite{prokofev04}. Then the only plausible scenario is that $\kappa>0$ 
throughout the glass phase (and there is no {\it a priori} reason to expect a very small $\kappa$). A key question then is whether $z=2$ or $z<2$. In the former case 
divergent SF clusters in the MI background close to the percolation point would be compressible, while in the latter case they should be incompressible.  
There are arguments for $z=2$ at the glass-SF transition \cite{fisher89} but no rigorous proofs. Some numerical works on models related to the BHM 
have in fact pointed to $z<2$ \cite{riyadarshee06,meier13}. Calculations suggesting $z=2$ are affected by large uncertainties 
\cite{krauth91,alet03,prokofev04} and are also consistent with $z<2$.

\begin{figure}
\center{\includegraphics[width=8.4cm, clip]{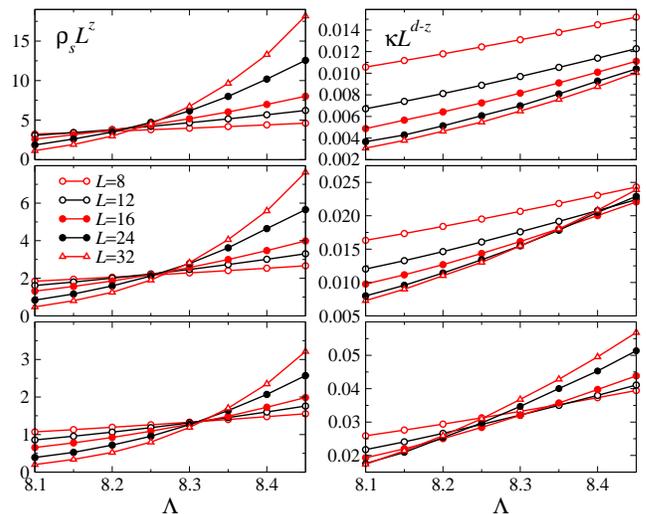}}
\caption{(Color online) Finite-size scaling of the superfluid stiffness (left column) and the compressibility (right column) 
for $U/t=22$ using three different values of the dynamic exponent; $z=2$, $1.75$, and $1.5$ (top to bottom) \cite{note}.}
\label{fss}
\end{figure}

We extract $z$ using the following finite-size scaling behaviors \cite{fisher89} expected exactly at the critical point:
\begin{equation}
\kappa (L,\Lambda_c) \propto L^{z-d},~~~~\rho_s (L,\Lambda_c) \propto L^{-z}.
\label{zscaling}
\end{equation}
In Fig.~\ref{fss} we show results using different system sizes and scaling the inverse temperature according to $\beta=L^z$ 
with three choices of $z$ \cite{note}. Based on Eqs.~(\ref{zscaling}) we expect curves of $\kappa L^{d-z}$ 
for different $L$ to cross each other at a point (asymptotically for large $L$) if the correct value of $z$ is used, and a similar behavior 
of $\rho_sL^z$. It should be noted, however, that there are crossings even if a wrong $z$ is used, but the vertical crossing value (e.g., for 
system sizes $L$ and $2L$) will then drift up or down instead of converging to a constant. One can also expect larger corrections to the horizontal 
crossing value if an  incorrect $z$ is used. The compressibility crossing points in Fig.~\ref{fss} are very sensitive to the value of $z$,
while the stiffness crossings are more stationary. Such behavior has also been observed in certain clean bosonic systems \cite{hohenadler11}.
Based on our result, $z$ should be between $1.5$ and $1.75$, which implies that the percolating SF cluster is incompressible.

In the above analysis it has been implicitly assumed that any non-singular contributions to $\kappa$ can be neglected. If regular contributions arise from SF 
domains larger than a critical size, then we would expect these contributions to increase with $L$ and, by Eq.~(\ref{zscaling}), this
would lead to an apparent {\it enhancement} of $z$. Since we instead find a reduction from $z=2$ it appears that non-singular background contributions 
are not responsible for this effect and $z<2$ should be a robust result. It is also interesting to note the distinct drop in compressibility in Fig.~\ref{l16} 
at the lower SF to glass transition, while at the second transition there are no strong variations, suggesting large regular contributions there, 
or, possibly, $z=2$.

Along with a large critical size needed for non-Mott domains to become superfluid, even far away from the Mott boundary, a dynamic
exponent $z < 2$ provides an explanation for an anomalously small, or possibly vanishing, $T=0$ compressibility in the finger region of the phase diagram
between the Mott and SF phases. The sharp cross-over from anomalously small to normal compressibility away from the SF phase at larger 
$U$ also shows that there are two distinct types of glass phases in the BHM, one being either an MG or an anomalous BG with physics similar to an MG, 
and the other one a standard BG with isolated non-coherent superfluid domains.

The scenario discussed here applies only to integer filling fractions, since a compressible state follows trivially from 
a non-constant $\rho(U,\Lambda)$ for incommensurable systems. Differences between integer and non-integer filling were found in a 
recent renormalization-group study \cite{kruger11}, although it is not clear whether the state is the MG or anomalous BG identified 
in our work. 

{\it Acknowledgments.}---We thank Claudio Chamon, Matthew Fisher, Philip Phillips, Anatoli Polkovnikov, Nikolay Prokof'ev, 
and Boris Svistunov for discussions. This research was supported by the NSFC under Grant No.~11175018 (WG) and by the NSF under Grants No.~DMR-1104708 and PHY-1211284 (AWS). 
WG would like to thank Boston University's Condensed Matter Theory Visitors program. AWS gratefully acknowledges travel support from
Beijing Normal University.

\appendix

\section*{Supplementary Material}

We here describe how we have determined the Mott gap of the clean Bose-Hubbard model, using the SSE QMC method and 
finite-size analysis. This gap allows for an accurate determination of the Mott--glass phase boundary of the site-disordered model.
We also discuss how the standard scenario of non-interacting superfluid ``lakes'' in the site-disordered model gives 
rise to a non-zero compressibility $\kappa$ in the limit $T\to0$. This scenario does not take into account finite-size 
effects of the lakes (among other approximations), and including those effects significantly reduce the value $\kappa (T\to 0)$.
We demonstrate this by studying a single domain with altered chemical potential inside a Mott insulator. To demonstrate 
sufficient statistics in the calculation of the disorder-averaged compressibility $\kappa$, we also present probability
distributions (histograms) of $\kappa$. They exhibit some tails in the glass phase, but not significant enough to cause
sampling problems related to rare occurrences at the temperatures considered. Finally, we demonstrate the small effects 
of the cut-off imposed on the site occupation numbers $n_i$ in the SSE simulations.

\subsection{Determination of the Mott gap}
\label{sec:gap}

It is not easy to determine the phase boundary between the Mott insulating (MI) and glass phase of the site-disordered
Bose-Hubbard model (BHM) by direct studies of properties of the glass phase, because in that region its properties are completely 
dominated by very large and extremely rare ``lakes'' of the non-Mott phase (which are normally assumed to be superfluid) inside the Mott 
background.\cite{freericks96} It is, however, well known that the phase boundary is completely determined by the Mott gap $\Delta_M$ 
of the clean system as it is this gap that has to be overcome by the average chemical potential of a large lake for such a lake to become 
superfluid (see recent discussion and calculations in, e.g., Refs.~\onlinecite{pollet09,gurarie09,soyler11}). It is easy 
to find the density as a function of chemical potential in a clean finite system by using QMC methods, and the Mott gap can be 
extracted from such curves as the width of the constant part, where the density $\rho$ is integer. Determining the MI--glass boundary 
by finite-size scaling of the gaps is the most straight-forward aspect of studies of the phase diagram of the site-disordered BHM 
in the plane $(U,\Lambda)$ of Coulomb repulsion and disorder strength.

As a prelude to discussing finite-size effects of the non-Mott lakes in Sec.~\ref{sec:lake}, we here demonstrate our calculations 
of the Mott gap with value quoted in the main text, including effects of the cut-off $n_{\rm max}$ of the site occupation numbers $n_i$ 
in the SSE QMC simulations.\cite{sse}

\begin{figure}
\includegraphics[scale=0.48]{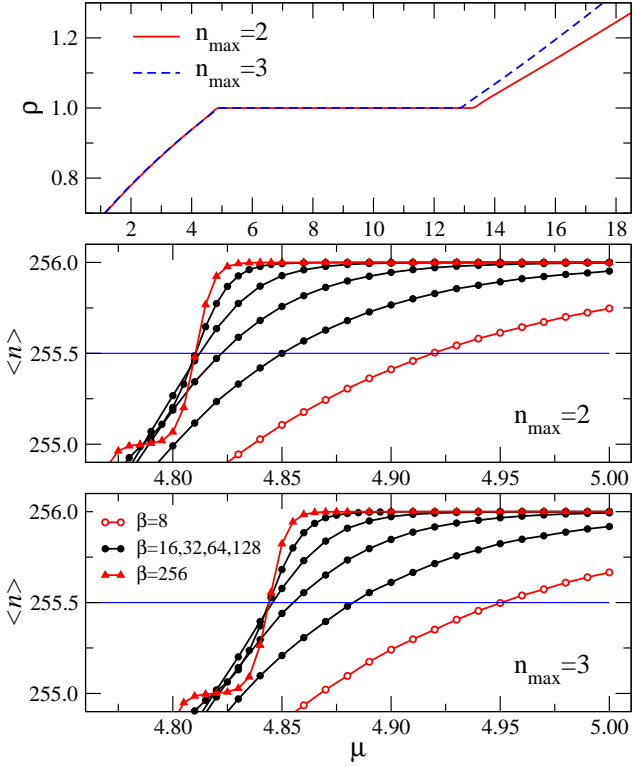}
\centering
\caption{(Color online) SSE QMC results for the particle density $\rho$ and the particle number $\langle n\rangle = N\rho$
of the clean $L=16$ BHM with $U=22$ as a function of the chemical potential $\mu$. Two different cut-offs, $n_{\rm max}=2,3$,
were used. The top panel shows results at inverse temperature $\beta=32$ in the whole region close to the $\rho=1$ plateau.
The middle ($n_{\rm max}=2$) and bottom ($n_{\rm max}=3$) panels show details around the left edge of the plateau at several 
inverse temperatures. The $\mu$-value at which the large4-$\beta$ curves cross the line $\langle n\rangle = N-1/2$ is used along 
with the analogous crossing with $\langle n\rangle = N+1/2$ at the right end of the plateau to extract the Mott gap (i.e.,
the width of the $\langle n\rangle = N$ plateau).} 
\label{kappa2}
\end{figure}

The density versus chemical potential is shown for a $16 \times 16$ system for $n_{\rm max}=2$ and $3$ in the top panel of Fig.~\ref{kappa2}.
The $\rho=1$ plateau is very clearly visible 
and slightly narrower for $n_{\rm max}=3$ (and naturally the effects of the cut-off are larger on the right-hand side of the plateau, where the 
particle number fluctuates to higher values). A large number of smaller density jumps for higher and lower chemical potentials are smeared 
into smooth curves by finite-temperature effects, even at the relatively low temperature ($\beta=1/T=32$) used here. The middle and bottom panels 
of Fig.~\ref{kappa2} show results at several different temperatures in the neighborhood of the left edge of the $\rho=1$ plateau. Here the 
particle number, $\langle n\rangle = N\rho$, is graphed instead of $\rho$. Plateaus at all integer values are expected in $\langle n\rangle$ 
versus $\mu$ if the temperature is low enough. Here we can clearly see the step between $\langle n\rangle = 255$ and $256$ developing for $\beta=256$. 
The $\mu$ point at which the curves cross the half-way distance between these two particle numbers provides a convenient and well convergent (versus
$\beta$) way of defining the edge of the $\rho=1$ plateau, which then is used in combination with the analogous step on the right-side 
to extract the Mott gap (here for the two different cut-offs).

\begin{figure}
\includegraphics[width=7.5cm]{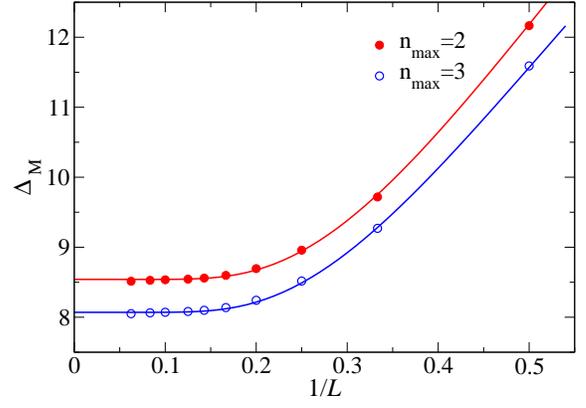}
\caption{(Color online) Determination of the Mott gap in the thermodynamic limit for the BHM with $\rho=1$ at $U=22$. The
points are results of SSE simulations such as those shown in Fig.~\ref{kappa2}, with particle-number cut-offs equal to $2$ and $3$, 
and the curves are fits to the form $\Delta_M(L)=\Delta_M(\infty) + a{\rm e}^{-bL}$, using system sizes $L \ge 5$. The results are
$\Delta_M=8.52(2)$ ($n_{\rm max}=2$) and $\Delta_M=8.10(2)$ ($n_{\rm max}=2$).}
\label{mottgap}
\end{figure}

In Fig.~\ref{mottgap} we show the finite-size scaling of $\Delta_M(L)$ extracted for several different system sizes at $U=22$. 
Given that the system has a finite correlation length, the gap is expected to converge exponentially fast, once $L$ exceeds this
correlation length, and our results are consistence with such a form. The extrapolated Mott gap is $\Delta_M=8.52(2)$ and
$\Delta_M=8.10(2)$ for $n_{\rm max}=2$ and $n_{\rm max}=3$, respectively. The latter value is very close to the value obtained 
in Ref.~\onlinecite{capogrosso08} without restrictions on the occupation numbers.

\subsection{Independent superfluid lakes scenario}

Consider a system of $N$ sites. Denoting the total number of bosons by $n$, the compressibility is given by
\be
{\kappa}=\frac{1}{NT} \left[\left< n^2\right>-\left< n \right>^2\right].
\label{kappa-def}
\ee
For a system in the Mott phase $n$ is an integer multiple of $N$ with no fluctuations at $T=0$.

Consider now a simplified model of a system with random potentials in which clusters of $N_i$ sites form lakes with chemical 
potential $\mu_i = \mu + \epsilon_i$, with the disorder values $\epsilon_i$ uniformly distributed in the interval $[-\Lambda,\Lambda]$ 
and the lakes labeled $i=1,\ldots,N_l$. Sites that do not belong to these clusters form a Mott-insulating background with chemical potential 
$\mu$ at the center of the Mott gap. Assume further that the Mott background is completely ``rigid'', so that the superfluid lakes 
can be considered completely independent of each other (i.e., the numbers $n_i$ of particles in each lake are uncorrelated). Then the
compressibility can be written as
\be
{\kappa}  = \frac{1}{NT} \sum_{i=1}^{N_l}\left[\left< n_i^2\right>-\left< n_i \right>^2\right] 
\label{kappa-def2}
\ee
We will here consider the case where all the sizes $N_i$ are finite, including also in the limit where the system size $N \to \infty$,
to illustrate the large contributions from finite lakes in the standard scenario.

\begin{figure}
\includegraphics[scale=0.48]{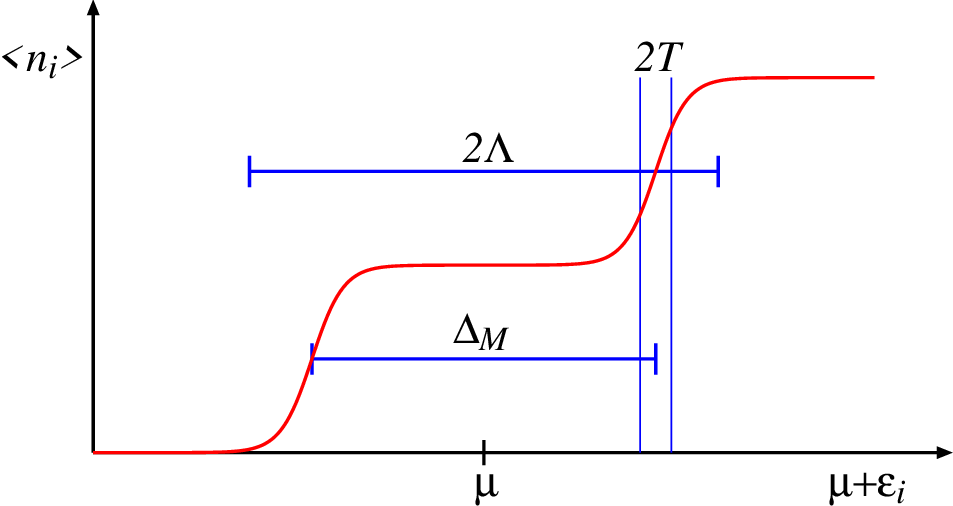}
\centering
\caption{(Color online) Particle number of a lake versus the total chemical potential $\mu + \epsilon_i$ of the lake. 
The width of the disorder distribution from which $\epsilon_i$ is drawn is $2\Lambda$. Lakes for which $\mu + \epsilon_i$ 
falls within a distance roughly equal to the temperature from the value of $\mu$ at which the particle number jumps 
at $T=0$ (this value being $\mu=\mu_0+\Delta_M/2$, where $\mu_0$ is the chemical potential at the center of the plateau)
contribute to the $T\to 0$ compressibility.}
\label{suppfig1}
\end{figure}

Figure \ref{suppfig1} schematically illustrates the expected dependence of the average particle number of one of
the lakes as a function of their total chemical potential (background value $\mu$ plus the random value $\epsilon_i$). 
The plateaus shown corresponds to $\rho=1$ (center) and one particle being added (right) or removed (left). If the half-width 
$\Lambda$ of the disorder distribution exceeds half the Mott gap, $\Lambda > \Delta_M/2$, then the lake can have particle 
density different from the Mott background. We will estimate the contribution to the $T \to 0$ compressibility from all the lakes 
in this case. Note that this estimation only includes contributions from each lake coming from fluctuations by $\pm 1$ particle. 
If $\Lambda$ is sufficiently large higher fluctuations also contribute but do no change the overall picture we want 
to illustrate.

Roughly, as indicated in Fig.~\ref{suppfig1}, there is a significant contribution from lake
$i$ if $\mu + \epsilon_i$ falls within a distance $T$ (or a few times $T$) from the step 
at which the particle number
of the lake changes. Within this window $\left< n_i^2\right>-\left< n_i \right>^2$ is of order
one, while outside the window this quantity is essentially zero. The probability of being
inside the window is $T/\Lambda$, and, thus, on average a lake contributes of the order 
$1/(N\Lambda)$ to the total compressibility (\ref{kappa-def2}). Since the number of lakes $N_l 
\propto N$  we have the standard result that the compressibility of a system of such lakes
with random chemical potentials is non-zero and not very small as $T \to 0$. Clearly this would
remain true also if the site energies are random within the lakes, in which case $\epsilon_i$ above
should be the average chemical potential of a lake. 

One crucial assumption made above was that $\Lambda$ was larger than half of the Mott gap,
so that particle-number sectors different from those corresponding to the Mott insulator can
be reached. However, if the lakes are finite one should not here use the value of the Mott gap in 
the thermodynamic limit, but the finite-size Mott gap corresponding to the lake size (and also there will
be shape effects). This size effect is likely at the heart of our finding that $\kappa$ of the site-disordered BHM 
is very small in an extended region, not only very close to the MI-glass boundary but also well away from it. 
The finite-size Mott gap of course not only depends on the size of the lake but also on its shape. In the next section 
we will illustrate the effect of finite size using a few simple cluster shapes. 

\subsection{Single lake in a Mott background}
\label{sec:lake}

Consider a lake A of arbitrary shape formed by a cluster of $N_A$ sites with chemical potential $\mu_A=\mu+\epsilon_A$ 
embedded in a Mott insulating background B with chemical potential $\mu_B=\mu$ and periodic boundary conditions.
The whole system has $N=L^2$ sites. The number of bosons in the system is $n$ with $n_A$ in A and $n_B$ in B. 

The compressibility of the whole system defined in (\ref{kappa-def}) can be expressed as 
\ba
{\kappa}
&=&\frac{1}{NT} \left[\left<\left(n_A+n_B\right)^2\right>
           -\left<n_A+n_B\right>^2\right].
\label{kdef2}
\ea
Consider $\mu_A$ not far away from $\mu_B$, so that the ground state of the whole system is in the Mott phase with $n$ an integer 
multiple of $N$. The total compressibility at $T=0$ vanishes. However, the number of bosons can still fluctuate in either subsystem by 
quantum tunneling through the boundaries. This example illustrates the fact that the basic assumption of the standard scenario and 
Eq. (\ref{kappa-def2}) are never strictly correct when $T \to 0$. The correlation between bosons in the lakes and in the background 
should be considered below some temperature which depends on the nature of the Mott background, the typical distance between the lakes, etc. 

Now consider temperatures higher than these tunneling-dominated low temperatures. Then the thermal energy dominates, leading to mainly 
independent grand-canonical number fluctuations within the lakes. If $\mu_A$ is turned up large enough, we expect $n=n_A+n_B$ at low $T$ to 
exhibit sharp (if $T$ is not too high) steps, where the system becomes compressible. It is clear that the completely rigid Mott background 
assumption of the standard scenario, Eq. (\ref{kappa-def2}), must to some extent fail at these compressible regions even at elevated temperatures. 
Strictly speaking it is the particle number $n$ of the total system that jumps, though most of the excess or deficit particle number will be 
localized inside the region $A$ of the modified chemical potential; the spread in excess density should be exponentially localized around 
region $A$. Thus, the boson number fluctuation inside lake A gives the essential contributions to the compressibility of the system when 
$\epsilon_A$ exceeds a certain value, which, according to the standard scenario, is half of the Mott gap. 
One can then apply the picture of independent lakes to the compressibility. 

To elucidate the finite-size effects in this picture, we have carried out illustrative calculations for a single domain with modified (variable) 
chemical potential in a background of fixed chemical potential. We focus on the following question: At which $\mu_A$ does the embedded domain start 
to contribute to the compressibility? To answer this question we use SSE QMC calculations to study finite clusters embedded in a larger MI system. 
We set the chemical potential of the MI background to $\mu_B=9.07$, which, as seen in Fig.~\ref{kappa-k2}, is at the center of the $\rho=1$ Mott 
lobe of the clean BHM. As it is not possible to define an area surrounding the domain which would completely single out the compressibility contribution 
of the domain alone, we investigate the compressibility of the full system, i.e., the fluctuation of the total particle number $\langle n\rangle$.
As discussed above, this should be essentially the same as the compressibility due to the embedded domain, given that the MI environment has 
a very substantial Mott gap and by itself has a very small compressibility.

\subsubsection{Finite-size effect for an embedded square cluster}

\begin{figure}
\bigskip
\includegraphics[scale=0.38]{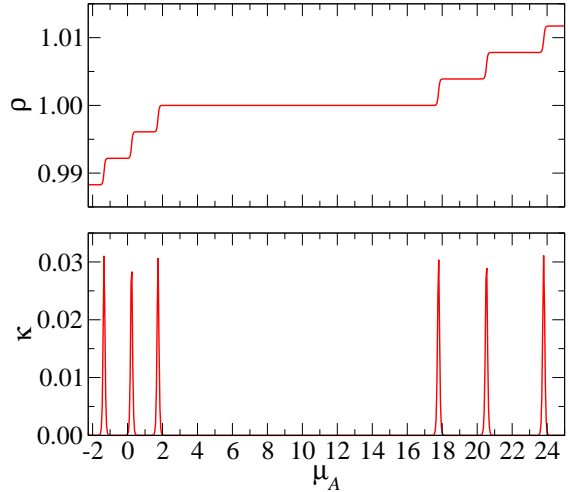}
\centering
\caption{(Color online) The particle density $\rho$ (top panel) and the 
compressibility $\kappa$ (bottom panel) of the $U=22$ system with $L=16$ at $\beta=32$ 
as functions of the chemical potential $\mu_A$ of the embedded lake formed by an $L_A=2$ cluster. }
\label{kappa-k2}
\end{figure}

\begin{figure}
\bigskip
\includegraphics[scale=0.38]{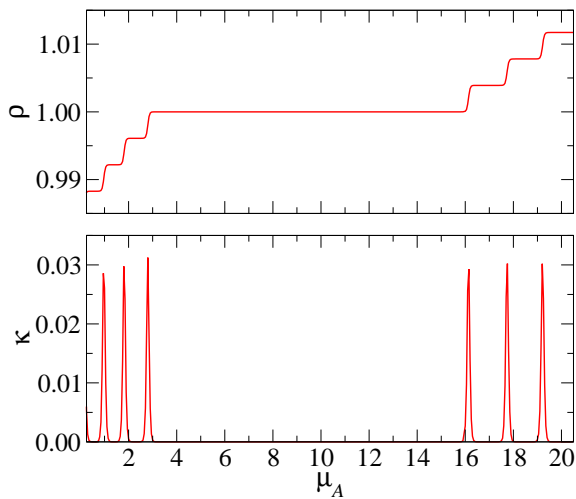}
\centering
\caption{(color online) The same quantities as in Fig.~\ref{kappa-k2} but with a lake-size $L_A=3$.}
\label{kappa-k3}
\end{figure}

We consider a system with $U=22$, studying first a square-shaped lake with linear size $L_A$ embedded in a system with $L=16$. Figures \ref{kappa-k2} 
and \ref{kappa-k3} show the density of bosons $\rho$ and the compressibility $\kappa$ as functions of $\mu_A$ of an $L_A=2$ and 
an $L_A=3$ lake, respectively. Here the inverse temperature is $\beta=32$, which is low enough to resolve clearly the 
different particle-number sectors of relevance to our discussion here.

The compressibility $\kappa$ is found to peak at values of $\mu_A$ corresponding to the steps in the density  
curves, where different particle-number sectors of the whole system are degenerate. The first peak on the right-hand side appears 
at $\mu_A(+)=17.69$ for $L_A=2$ and at $\mu_A(+)=16.07$ for $L_A=3$, respectively. These values correspond to the deviation in
chemical potential,
\begin{equation}
\epsilon_A(+)=\mu_A(+)-\mu_B,
\end{equation}
being much larger than $\Delta_M/2 \approx 4.3$. The first peak on the left-hand side appears at $\mu_A(-)=1.69$ 
for $L_A=2$ and at $\mu_A(-)=2.77$ for $L_A=3$, respectively. Again, they correspond to the deviation
\begin{equation}
\epsilon_A(-)= \mu_B-\mu_A(-)
\end{equation}
being much larger than $\Delta_M/2$. 

\begin{figure}
\bigskip
\includegraphics[scale=0.44]{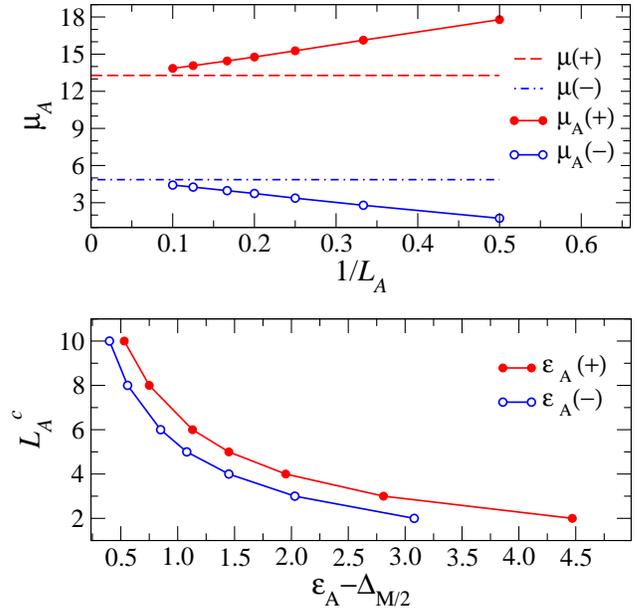}
\centering
\caption{(Color online) The top panel shows the finite-size Mott-gap boundaries $\mu_A(\pm)$ for a square cluster embedded 
in a Mott background as functions of the cluster size $L_A$. The boundaries are compared with the infinite-size Mott boundaries 
$\mu(+)-\mu(-)=\Delta_M$. The bottom panel shows the characteristic size $L_A^c$ beyond which an embedded cluster with chemical-potential
shift $\epsilon_A$ leads to a compressible system.}
\vskip-2mm
\label{muc-L}
\end{figure}

The value $\mu_A(+)-\mu_A(-)$ can be understood as an effective Mott gap of the finite embedded cluster, which is much larger than 
the Mott gap $\Delta_M$. As the cluster size $L_A$ increases, the gap shrinks and should approach $\Delta_M$ in the limit 
$L_A \to \infty$, as it appears to do roughly in the top panel of Fig.~\ref{muc-L}, where results are graphed for several $L_A$. There
will of course be secondary finite-size effects here as well (i.e., beyond the dependence on $L_A$) when $L_A$ approaches the full-system 
size $L$.

The above results imply that, given $\epsilon_A(\pm) >\Delta_M/2$ of a lake embedded in the MI, to contribute to the system's
compressibility the cluster has to be larger than a characteristic size $L_A^c$. This size extracted from the data in the upper
panel of Fig.~\ref{muc-L} are graphed in the bottom panel of Fig.~\ref{muc-L}, for both added and removed particles. Clusters with size 
smaller than $L_A^c$ are actually still Mott-like and should not be considered as superfluid lakes (though in general their properties 
can still be distinct from the Mott background, due to their altered spectral properties, which are related to their sizes and shapes). 
In the limit $\epsilon_A(\pm) \to \Delta_M/2$, the embedded cluster has to be infinitely large to contribute to the compressibility and
this explains why the compressibility must still be exponentially decaying with temperature right at the MI--glass boundary,
as we found in the main text.

\subsubsection{Effect of cluster shape}

\begin{figure}
\includegraphics[scale=0.38]{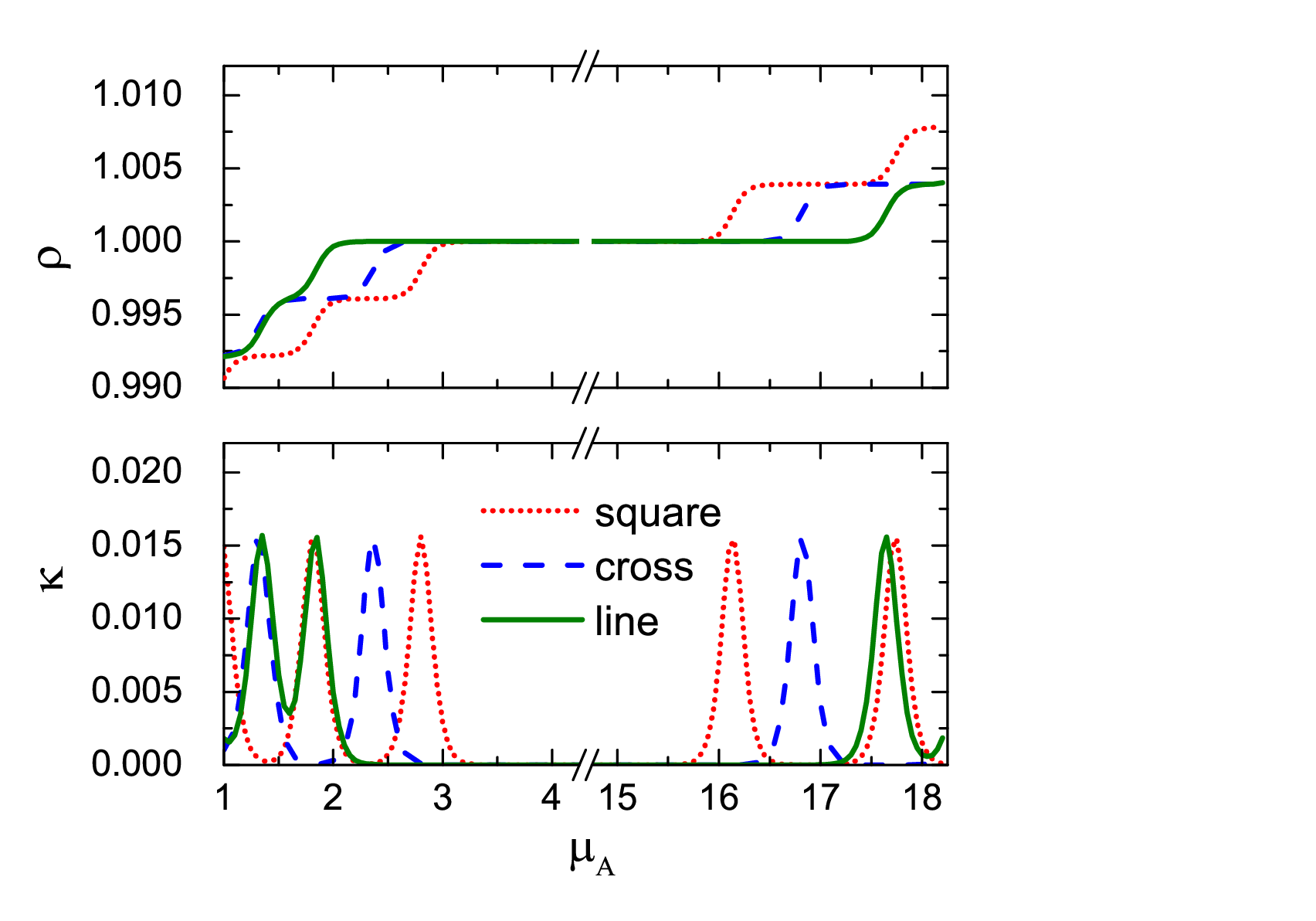}
\vskip-4mm
\caption{(Color online). The density $\rho$ (top panel) and compressibility $\kappa$ (bottom panel) versus
the chemical potential $\mu_A$ of the embedded clusters in the square, cross and line-like shape, respectively. In
all cases the cluster size is $9$ sites and the inverse temperature $\beta=16$.}
\vskip-1mm
\label{shape}
\end{figure}

The embedded clusters are in reality, in an actual site-disordered BHM, typically not of square shape. We now consider effects 
of cluster shape by comparing embedded clusters in the shape of a square, a cross, and a line. The number of sites in the clusters, 
embedded in the $L=16$ Mott background, is $N_A=9$ in all cases. The inverse temperature in the SSE simulations was $\beta=16$,
leading to significant temperature smoothing but still resolvable steps in the particle number, as shown in the top panel of Fig. \ref{shape}.
The line-shaped cluster and cross-shaped clusters have larger effective Mott gaps than that of the square cluster. This means that, for a 
given deviation $\epsilon_A(\pm)>\Delta_M/2$, the characteristic size required for non-zero compressibility of an irregularly shaped cluster is even 
larger than that of the squared one discussed in the previous section.

\subsubsection{Conclusions}

The main point argued here is that the Mott gap of ``lake'' embedded in the Mott background is enhanced by finite size (of the lake), 
and, as a direct consequence of this, the compressibility of the glass phase
can be much smaller than would be expected based on the standard scenario of superfluid lakes (as most of the lakes will actually
not contribute significantly). As discussed in the 
main text, effects of quantum-criticality with dynamic exponent $z \not=2$ can further contribute to an anomalously 
small compressibility also close to the superfluid phase.

While we did not discuss in details effects of correlations in particle-number fluctuations between different lakes, 
these can also be expected to be ultimately important in the limit $T \to 0$. It would be interesting to carry out 
detailed studies of the effects of such correlations by considering two or more domains with altered chemical potential 
in a Mott background, generalizing the single-lake calculations presented here.

\subsection{Compressibility distribution}

\begin{figure}
\includegraphics[scale=0.48]{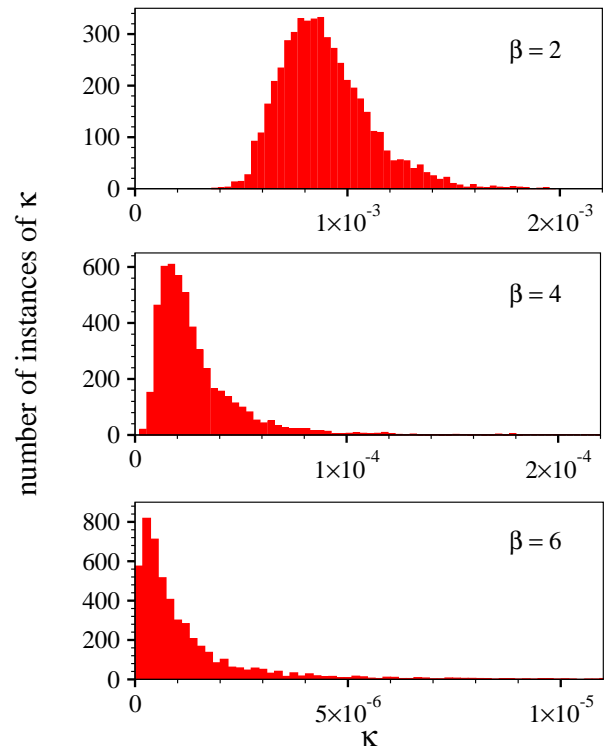}
\caption{(Color online). Histograms of the compressibility computed using approximately 5000 realization of random
site potentials for an $L=16$ lattice at $U=22$, $\Lambda=6$, and different values of the inverse temperature (as indicated
in the panels)}.
\vskip-4mm
\label{hist}
\end{figure}

When considering disorder averages there is always a potential concern of long tails in the probability distributions
of computed quantities, which may not be adequately samples with the relatively small number of disorder realizations 
used in typical QMC simulations. This issue may be particularly worrisome in a glass phase, where the low-energy fluctuations 
can be completely dominated by very rare configurations---in this case large lakes of superfluid in the MI background. If the 
anomalous glass phase we have demonstrated has a non-zero but very small compressibility in the limit $T \to 0$ [$c>0$ in Eq.~(2) in the
main text], that very small compressibility should be exactly due to rare large superfluid lakes. Then the stretched exponential form we 
found for the $T$-dependent compressibility in the main text should exhibit a cross-over to a small essentially $T$-independent 
value below some temperature. We have not seen any signs of such cross-overs in the ``finger'' region in the phase diagram, but one 
may worry that the unexpectedly small compressibility and absence of a cross-over could be due to poor sampling of rare
disorder realizations with large compressibility. To investigate this issue, we have constructed histograms approximating
the probability distribution $P(\kappa)$, examples of which are displayed in Fig.~\ref{hist}. While one may argue that very rare large-$\kappa$ 
``events'' are not sampled in practice, the probability distributions must have well-behaved functional forms, and one should be able to
estimate effects of large-$\kappa$ regions not sampled by investigating the tail-parts of the distributions that actually 
are sampled in simulations with a typical number (here thousands) of random samples.

It is clear, as seen in Fig.~\ref{hist}, that the distribution becomes increasingly skewed as the temperature is lowered, with a thin high-$\kappa$ tail 
developing. The whole distribution moves rapidly toward lower values, however, and the tail is not sufficiently long or fat to cause serious problems 
with properly converging the mean compressibility. The results are consistent with an exponentially decaying tail and all indications are that 
we have collected an adequate number of configurations for correctly estimating the disorder-average of the compressibility at all 
temperatures studied. We can of course still not rule out more significant tails at lower temperatures than we have studied here.
Relatively longer and more prominent tails in fact appear likely based on our results, but do not by themselves necessarily imply a non-zero 
$\kappa (T \to 0)$. 

\subsection{Effects of the particle-number cut-off }

In the SSE method we have used,\cite{sse} it is necessary to impose a cut-off on the site-occupation numbers, $n_i \le n_{\rm max}$. This can be 
seen as an approximation, or it can simply be regarded as modifying the BHM with an effective $U$ which depends on the occupation 
number, i.e., $U(n_i)=\infty$ for $n_i > n_{\rm max}$. This is a perfectly reasonable model with the same symmetries as the original
BHM. In practice, the BHM with large $U$ ``self-imposes'' a cut-off, in the sense that the probability of occupying a site 
with many particles becomes exceedingly small. In the SSE method one can also use a high enough cut-off for there to be no detectable 
effects of the cut-off at all. However, since we are studying disordered systems and have to carry out averages over (typically) thousands of 
realizations of the random potentials, it is useful to have a small cut-off because that speeds up the calculations. 
The salient properties of these two versions of the model should not be expected to differ much in the parameter regions we have studied.

\begin{figure}[t]
\includegraphics[scale=0.32, clip]{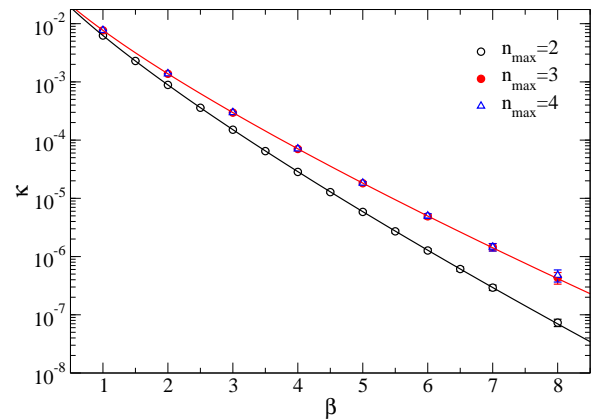}
\centering
\caption{(Color online) The compressibility at $U=22$, $\Lambda=6$ computed with different cut-offs $n_{\rm max}$ in
SSE simulations of $L=16$ systems. The curves are fits to the stretched exponential form, Eq.~(\ref{kappaexpon}),
using $\beta=4$ and higher in the fits (but also the behavior at lower $\beta$ is seen to match well the resulting fits).
The parameters are $b\approx 2.85, \alpha \approx 0.78$ for $n_{\rm max}=2$ and $b\approx 2.66, \alpha \approx 0.75$ 
for $n_{\rm max}=3$ (and there are no statistically significant differences between $n_{\rm max}=3$ and $4$).}
\label{nmax-L16}
\end{figure}

\begin{figure}[t]
\includegraphics[scale=0.32, clip]{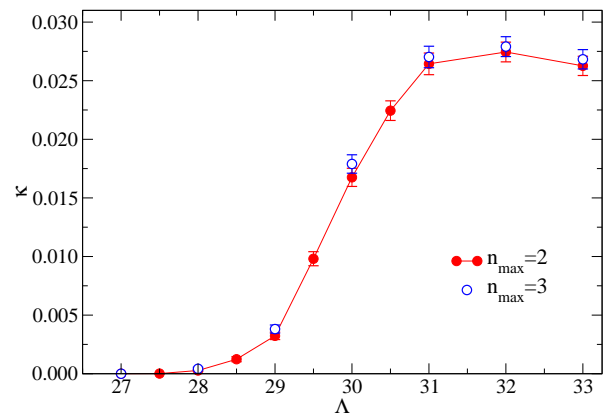}
\centering
\caption{(Color online) Compressibility versus disorder strength for $L=8$ systems at $U=60$ and inverse temperature $\beta=8$, 
calculated with cut-offs $n_{\rm max}=2$ and $3$.}
\label{U60-L8}
\end{figure}

Nevertheless, to check for effects of the cut-off, in addition to the main 
results obtained with $n_i \le 2$, we have also carried out some calculations with $n_i \le 3$ and $n_i \le 4$. We have already discussed 
the small effects of restricting the on-site occupation numbers to $n_i \le n_{\rm max}$ in Sec.~\ref{sec:gap}. Here we discuss effects 
on the compressibility.

Fig.~\ref{nmax-L16} shows the $\Lambda=6$ data from Fig.~3 in the main text, which was computed with $n_{\rm max}=2$, along with data
also for $n_{\rm max}=3$ and $4$. While, as would be expected, a higher cut-off enhances the compressibility, the stretched exponential
form proposed in the main text,
\begin{equation}
\kappa \propto {\rm e}^{-b/T^\alpha}
\label{kappaexpon}
\end{equation}
is robust. Interestingly, the exponent $\alpha$ only changes from $\alpha \approx 0.78$ for 
$n_{\rm max}=2$ to $\alpha \approx 0.75$ for $n_{\rm max}=3$, while the effect on the prefactor $b$ multiplying 
$T^{-\alpha}$ in the exponential
is somewhat larger (values given in the caption of Fig.~\ref{nmax-L16}). Within the temperature range studied, there are no detectable 
differences between the results for $n_{\rm max}=3$ and $n_{\rm max}=4$.

Fig.~\ref{U60-L8} shows the compressibility in the parameter region of the inset of Fig.~2 in the main paper, but using $L=8$ 
instead of $L=16$ and comparing results for $n_{\rm max}=2$ and $3$. There are no significant differences
between the two data sets (perhaps just a small enhancement of $\kappa$ when $n_{\rm max}=3$). In addition to confirming the 
negligible effects of the cut-off $n_{\rm max}=2$, these results also demonstrate the small size effects in this 
region (as the results agree very closely with those for $L=16$ in the main paper). The small effects of the cut-off in this region 
is perhaps not surprising considering that the repulsion $(U/2)n(n-1)=180$ for $n=3$, making triple and quadruple occupancy of 
the sites very small even on sites with the smallest site energies even when $\Lambda \approx 30$.

In conclusion, the restriction $n_i \le n_{\rm max}=2$ considered in the main paper some times has small quantitative effects on the
results (relative to having no cut-off at all), but qualitatively the properties of the model are the same, e.g., a stretched
exponential form of the temperature dependent compressibility in the state we have termed the anomalous Bose-glass, and the
existence of a sharp cross-over (perhaps turning into a phase transition in some parts of the phase diagram) between this almost 
incompressible state and a normal highly compressible Bose-glass. One would also not expect universal physics of the model to
depend on the cut-off (e.g., the dynamic exponent $z$ determined in the main text), as no symmetries are changed and the cut-off 
also can be considered as a minor modification of the on-site repulsion.

\null

\end{document}